\documentclass[final,5p,twocolumn,times,sort&
compress]{elsarticle}

\usepackage{mathrsfs}
\usepackage[latin1]{inputenc}
\usepackage{array}
\usepackage{dsfont}
\usepackage{amsmath}\usepackage{amssymb,stmaryrd}
\usepackage{booktabs}
\usepackage{lipsum}
\usepackage{cancel}
\usepackage{xcolor}
\usepackage{hyperref}
\usepackage[dvips]{feynmp}
\def\preprintdate{May 2019}

\renewcommand{\eqref}[1]{\mbox{Eq.~(\ref{#1})}}

\newcommand{\secref}[1]{\mbox{Sec.~\ref{#1}}}

\begin{document}

\begin{frontmatter}

\title{Dimensional reduction of the electromagnetic sector \\ of the nonminimal Standard-Model-Extension}

\author{Manoel M. Ferreira Jr., Jo\~{a}o A.A.S. Reis, and Marco Schreck}

\address{Departamento de F\'{\i}sica, Universidade Federal do Maranh\~{a}o (UFMA) \\
Campus Universit\'{a}rio do Bacanga, S\~{a}o Lu\'{\i}s -- MA, 65080-805, Brazil}

\address{manojr.ufma@gmail.com, jalfieres@gmail.com, marco.schreck@ufma.br}

\address{\rm
\preprintdate
}

\begin{abstract}

In the current paper, we construct a Lorentz-violating electrodynamics in (1+2) spacetime dimensions from the electromagnetic sector of the nonminimal Standard-Model Extension (SME) in (1+3) dimensions. Subsequently, we study some of the basic properties of this framework. We obtain the field equations, the Green's functions, and the perturbative Feynman rules. Furthermore, the modified dispersion relations are computed at leading order in Lorentz violation. We then remove the unphysical degrees of freedom from the electromagnetic Green's function that are present due to gauge invariance. The resulting object is used to construct the general solutions of the uncoupled field equations with external inhomogeneities present. This modified planar electrodynamics may be valuable to describe electromagnetic phenomena in two-dimensional condensed-matter systems. Furthermore, it supports a better understanding of the electromagnetic sector of the nonminimal SME.

\end{abstract}

\end{frontmatter}

\section{Introduction}

Violations of Lorentz invariance have been hypothesized to emerge from phenomena related to Planck-scale physics such as strings~\cite{Kostelecky:1988zi}, spin networks described by loop quantum gravity~\cite{Gambini:1998it}, noncommutative spacetimes~\cite{AmelinoCamelia:1999pm}, spacetime foam~\cite{Klinkhamer:2003ec}, nontrivial spacetime topologies~\cite{Klinkhamer:1998fa}, and effects connected to UV completions of general relativity with Ho\v{r}ava-Lifshitz gravity~\cite{Horava:2009uw} as a prominent example. The Standard-Model Extension (SME)~\cite{Colladay:1996iz} is a well-established framework to parameterize possible deviations from Lorentz invariance in \textit{vacuo}. It is an effective field theory in (1+3) spacetime dimensions including the Standard-Model fields and the Riemann curvature tensor when gravity is taken into account.

Each Lorentz-violating contribution is decomposed into a background field and a field operator that the background is contracted with. The Lagrange density of the SME transforms as a Lorentz scalar under coordinate changes (observer Lorentz transformations). However, the background fields transform trivially under Lorentz transformations of an experimental apparatus (particle Lorentz transformations), whereupon the theory is not invariant under transformations of this kind. Each background field involves preferred spacetime directions and controlling coefficients describing the magnitude of Lorentz violation. At the level of effective field theory, a violation of discrete \textit{CPT} symmetry implies a violation of Lorentz invariance~\cite{Greenberg:2002uu}. Therefore, a subset of the contributions in the SME is \textit{CPT}-violating.

The SME was initially constructed to include field operators of mass dimensions 3 and 4 \cite{Colladay:1996iz}. The latter framework is called the minimal SME and it contains a finite number of controlling coefficients. To date, quite a large number of these coefficients have been tightly constrained by experimental tests~\cite{Kostelecky:2008ts}. Therefore, there was the need to generalize the minimal SME to the nonminimal SME \cite{Kostelecky:2011gq,Kostelecky:2009zp}. The latter contains an infinite amount of controlling coefficients that are contracted with field operators of arbitrary mass dimension $d$. The mass dimension of these field operators is increased by including additional four-derivatives. The more four-derivatives are present in a certain term, the more dominant this particular term becomes with increasing energy.

Contributions of the nonminimal SME were shown to be generated in noncommutative field theories after applying the Seiberg-Witten map \cite{Carroll:2001ws}. They can also arise from quantum corrections when nonminimal couplings between fermions and photons are present \cite{Borges:2016} as well as within supersymmetric scenarios \cite{Bonetti:2017}. For the past years there has been a certain interest in improving our understanding of such nonminimal contributions. For example, certain operators that are part of the nonminimal electromagnetic sector of the SME were proposed and studied in \cite{Myers:2003,Ferreira:2019lpu}. Apart from that, operators of the nonminimal SME fermion sector were subject to investigations, as well~\cite{Schreck:2014fm}.

Recently, interest has arisen to describe condensed-matter systems based on the SME. Many interesting phenomena in condensed-matter physics occur in planar systems such as the quantum Hall effect~\cite{Klitzing:2017} or ``relativistic effects'' in graphene~\cite{Allen:2010}. In this sense, studying and applying a planar electrodynamics to real physical systems can be seen as a suitable and sound investigation proposal. One possibility of obtaining planar theories from parent models defined in (1+3) spacetime dimensions is by means of the procedure known as dimensional reduction. This method has been used before to derive the planar version of the Maxwell-Carroll-Field-Jackiw Lagrangian~\cite{Belich:2002vd}. It was also employed to obtain the planar version of the {\em CPT}-even electromagnetic sector of the SME~\cite{Manojr:2011vd}.

In the current work, we propose a framework of an electromagnetism modified by nonminimal Lorentz violation in (1+2) spacetime dimensions. To perform an analysis as general as possible, we choose the broad electromagnetic sector of the nonminimal SME \cite{Kostelecky:2008ts} as the parent theory. Our goal is to provide a nonminimal daughter electrodynamics in (1+2) dimensions by using dimensional reduction~\cite{Belich:2002vd,Manojr:2011vd}. The purpose of such a procedure is two-fold. First, a planar Lorentz-violating electrodynamics can be valuable to describe condensed-matter phenomena that occur in two-dimensional systems. Second, a treatment of a (1+2)-dimensional field theory is expected to be simpler than that of a (1+3)-dimensional one from a technical perspective. Therefore, gaining understanding of the (1+2)-dimensional field theory might even have impact on the parent theory.

The paper is organized as follows. In \secref{sec:dimensions-reduction} we apply dimensional reduction to the electromagnetic sector of the SME. In this context as well as in \secref{sec:connection-physical-fields} we discuss some basic properties of the resulting planar theory. These introductory considerations are followed by \secref{sec:field-equations} where the system of coupled field equations is obtained, in general, as well as the modified planar Maxwell equations, in particular. The Green's functions for the scalar and electromagnetic field are computed in \secref{sec:perturbative-expansion}. Here, we also determine the perturbative Feynman rules. Section~\ref{sec:modified-dispersion-relations} is dedicated to obtaining the dispersion relations of the scalar and the  electromagnetic field at leading order in Lorentz violation. In \secref{sec:solutions-free-field-equations}, the previous results are employed to eliminate the unphysical degrees of freedom from the Green's function of the electromagnetic field. The resulting physical Green's function serves as a base to construct the inhomogeneous solutions of the Maxwell equations in the presence of an external four-current. The inhomogeneous solutions of the scalar field equations are derived in an analog way, whereas in this case it is not necessary to get rid of unphysical modes. Finally, the findings are summarized and concluded on in \secref{sec:conclusions}. Technical details that may be of interest to some readers are presented in~\ref{sec:general-photon-propagator}. Natural units are used with $\hbar=c=1$ unless otherwise stated. Lorentz indices are denoted by Greek letters, whereas we indicate spatial indices by Latin letters.

\section{Dimensional reduction}
\label{sec:dimensions-reduction}

The present work is based on the electromagnetic sector of the nonminimal SME. In \cite{Kostelecky:2009zp} it is formulated via the Lagrange density
\begin{align}
\label{eq:sme-photon-sector}
\mathcal{L}_{\left(1+3\right)}&=-\frac{1}{4}F_{\hat{\mu}\hat{\nu}}F^{\hat{\mu}\hat{\nu}}+\frac{1}{2}\varepsilon^{\hat{\lambda}\hat{\kappa}\hat{\mu}\hat{\nu}}A_{\hat{\lambda}}(\hat{k}_{AF}) _{\hat{\kappa}}F_{\hat{\mu}\hat{\nu}} \notag \\
&\phantom{{}={}}-\frac{1}{4}F_{\hat{\kappa}\hat{\lambda}}(\hat{k}_{F})^{\hat{\kappa}\hat{\lambda}\hat{\mu}\hat{\nu}}F_{\hat{\mu}\hat{\nu}}\,,
\end{align}
where $A_{\hat{\mu}}$ is the \textit{U}(1) gauge field and $F_{\hat{\mu}\hat{\nu}}=\partial_{\hat{\mu}}A_{\hat{\nu}}-\partial_{\hat{\nu}}A_{\hat{\mu}}$ the field strength tensor associated. All fields are defined in (1+3)-dimensional Minkowski spacetime endowed with metric tensor $(\eta_{\hat{\mu}\hat{\nu}})=\mathrm{diag}(1,-1,-1,-1)$. Lorentz indices with hats refer to this spacetime, i.e., $\hat{\mu}\in \{0\dots 3\}$. Furthermore, $\varepsilon^{\hat{\lambda}\hat{\kappa}\hat{\mu}\hat{\nu}}$ is the Levi-Civita symbol in (1+3)  dimensions where we use the convention $\varepsilon^{0123}=1$.

The first term in \eqref{eq:sme-photon-sector} is the standard Maxwell term. The second is a \textit{CPT}-odd extension of the electromagnetic sector where $(\hat{k}_{AF})_{\hat{\kappa}}$ transforms as a four-vector under coordinate changes. The third is \textit{CPT}-even and includes the fourth-rank tensor $(\hat{k}_{F})^{\hat{\kappa}\hat{\lambda}\hat{\mu}\hat{\nu}}$. The objects $\hat{k}_{AF}$ and $\hat{k}_F$ are interpreted as sets of scalars under Lorentz transformations of an experimental apparatus. They are background fields that permeate the vacuum and give rise to preferred spacetime directions. In the minimal SME, they are introduced as controlling coefficients independent of the spacetime coordinates. This assumption is usually taken to guarantee that translation invariance is still a symmetry of the theory, which implies energy-momentum conservation. In  this context, the second contribution of \eqref{eq:sme-photon-sector} is denoted as Maxwell-Chern-Simons (MCS) term and the third is sometimes called modified Maxwell term.

In the nonminimal SME, the minimal controlling coefficients are promoted to operators that include additional four-derivatives. Now each of these operators is an infinite sum over controlling coefficients contracted with a number of four-derivatives that successively increases by 2. We will use the terms \textit{minimal} and \textit{nonminimal} in the same context within the modified planar electrodynamics to be constructed.

There are several procedures to derive a field theory of a (1+2)-dimensional electromagnetism from a (1+3)-dimensional parent theory. A first method could be a simple projection, i.e., to set the third component of the gauge field to zero and to disregard any dependence on the third spatial coordinate, e.g., $A^{\hat{\mu}}(t,\mathbf{x}^{(3)})\mapsto A^{\mu}(t,\mathbf{x}^{(2)})$ with $\mu\in \{0\dots 2\}$. Indices without a hat refer to (1+2)-dimensional Minkowski spacetime with metric tensor $(\eta_{\mu\nu})=\mathrm{diag}(1,-1,-1)$. As a short-hand notation, we introduce the spatial coordinates $\mathbf{x}^{(2)}\equiv (x,y)$ and $\mathbf{x}^{(3)}\equiv (x,y,z)$ for two and three spatial dimensions, respectively.

An alternative, more sophisticated approach to construct a (1+2)-dimensional daughter theory from a (1+3)-dimensional parent theory, is to disconnect the third component of $A^{\hat{\mu}}$ from the gauge field and to reinterpret it as a scalar field $\phi$ where the third spatial coordinate is again omitted:
\begin{subequations}
\begin{align}
A^{\hat{\mu}\neq 3}(t,\mathbf{x}^{(3)})&\mapsto A^{\mu}(t,\mathbf{x}^{(2)})\,, \displaybreak[0]\\[2ex]
A_{\hat{\mu}\neq 3}(t,\mathbf{x}^{(3)})&\mapsto A_{\mu}(t,\mathbf{x}^{(2)})\,, \displaybreak[0]\\[2ex]
A^{\hat{3}}(t,\mathbf{x}^{(3)})&\mapsto \phi(t,\mathbf{x}^{(2)})\,, \displaybreak[0]\\[2ex]
A_{\hat{3}}(t,\mathbf{x}^{(3)})&\mapsto -\phi(t,\mathbf{x}^{(2)})\,.
\end{align}
\end{subequations}
This technique is sometimes called dimensional reduction in the literature~\cite{Belich:2002vd}. Its advantage is that it automatically includes the first method simply for the choice $\phi=0$. Thus, this technique naturally gives rise to a scalar in contrast to simply putting it in by hand. Due to emergent couplings between the vector field in (1+2) dimensions and the scalar field, a plethora of additional interesting effects can emerge. If the scalar field is not suitable to describe a physical system, it can always be set to zero.

Because of the presence of Lorentz violation, dimensional reduction must also be applied to the background fields and the Levi-Civita symbol. In general,
\begin{subequations}
\begin{align}
(\hat{k}_{AF})^{\hat{\kappa}\neq 3}(t,\mathbf{x}^{(3)})&\mapsto (\hat{k}_{AF})^{\kappa}(t,\mathbf{x}^{(2)})\,, \displaybreak[0]\\[2ex]
(\hat{k}_{AF})^{\hat{3}}(t,\mathbf{x}^{(3)})&\mapsto \hat{k}_{AF}(t,\mathbf{x}^{(2)})\,, \displaybreak[0]\\[2ex]
(\hat{k}_{AF})_{\hat{\kappa}\neq 3}(t,\mathbf{x}^{(3)})&\mapsto (\hat{k}_{AF})_{\kappa}(t,\mathbf{x}^{(2)})\,, \displaybreak[0]\\[2ex]
(\hat{k}_{AF})_{\hat{3}}(t,\mathbf{x}^{(3)})&\mapsto -\hat{k}_{AF}(t,\mathbf{x}^{(2)})\,, \displaybreak[0]\\[2ex]
\varepsilon^{\lambda\mu\nu 3}&\mapsto \varepsilon^{\lambda\mu\nu}\,,
\end{align}
\end{subequations}
where $\varepsilon^{\lambda\mu\nu}$ is the Levi-Civita symbol in (1+2) dimensions. Analog correspondences can be established for the observer tensor $(k_F)^{\hat{\mu}\hat{\nu}\hat{\varrho}\hat{\sigma}}$. Even though the coefficients above have been written as functions of the spacetime coordinates, we will omit such dependencies to avoid a loss of translation invariance in the planar theory.

Carrying out this procedure for the individual terms of the Lagrange density (\ref{eq:sme-photon-sector}), which are contained in the nonminimal electromagnetic sector of the SME, leads to:
\begin{subequations}
\begin{align}
-\frac{1}{4}F_{\hat{\mu}\hat{\nu}}F^{\hat{\mu}\hat{\nu}}&\mapsto -\frac{1}{4}F_{\mu\nu}F^{\mu\nu}+\frac{1}{2}\partial_{\mu}\phi\partial^{\mu}\phi\,, \displaybreak[0]\\[2ex]
-\frac{1}{4}F_{\hat{\kappa}\hat{\lambda}}(\hat{k}_F)^{\hat{\kappa}\hat{\lambda}\hat{\mu}\hat{\nu}}F_{\hat{\mu}\hat{\nu}}&\mapsto-\frac{1}{4}F_{\kappa\lambda}(\hat{k}_{F})^{\kappa\lambda\mu\nu}F_{\mu\nu}-\partial_{\kappa}\phi(\hat{k}_{\phi})^{\kappa\mu}\partial_{\mu}\phi \notag \\
&\phantom{{}={}}+F_{\kappa\lambda}(\hat{k}_{\phi F})^{\kappa\lambda\mu}\partial_{\mu}\phi\,, \displaybreak[0]\\[2ex]
\frac{1}{2}\varepsilon^{\hat{\lambda}\hat{\kappa}\hat{\mu}\hat{\nu}}A_{\hat{\lambda}}(\hat{k}_{AF})_{\hat{\kappa}}F_{\hat{\mu}\hat{\nu}}&\mapsto -\varepsilon^{\lambda\kappa\mu}A_{\lambda}(\hat{k}_{AF})_{\kappa}\partial_{\mu}\phi-\frac{1}{2}\varepsilon^{\lambda\mu\nu}A_{\lambda}(\hat{k}_{AF})F_{\mu\nu} \notag \\
&\phantom{{}={}}-\varepsilon^{\mu\kappa\nu}\phi(\hat{k}_{AF})_{\kappa}\partial_{\mu}A_{\nu}\,,
\end{align}
\end{subequations}
where we defined a set of new operators via $(\hat{k}_{\phi})^{\kappa\mu}\equiv (\hat{k}_F)^{\kappa 3\mu 3}$ and $(\hat{k}_{\phi F})^{\kappa\lambda\mu}\equiv (\hat{k}_F)^{\kappa\lambda\mu 3}$. Here we see that only the \textit{CPT}-even term in (1+3) dimensions can generate a pure scalar contribution, whereas the \textit{CPT}-odd term does not do so. The planar Lagrange density obtained after dimensional reduction has the form
\begin{align}
\label{eq:lagrange-density-dimensionally-reduced}
\mathcal{L}_{(1+2)}&=-\frac{1}{4}F_{\mu\nu}F^{\mu\nu}+\frac{1}{2}\partial_{\mu}\phi\partial^{\mu}\phi-\frac{1}{2}\varepsilon^{\lambda\mu\nu}A_{\lambda}(\hat{k}_{AF})F_{\mu\nu} \notag \\
&\phantom{{}={}}-\frac{1}{4}F_{\kappa\lambda}(\hat{k}_{F})^{\kappa\lambda\mu\nu}F_{\mu\nu}-\partial_{\kappa}\phi(\hat{k}_{\phi})^{\kappa\mu}\partial_{\mu}\phi \notag \\
&\phantom{{}={}}+\varepsilon^{\nu\kappa\mu}\left[\phi(\hat{k}_{AF})_{\kappa}\partial_{\mu}A_{\nu}-A_{\nu}(\hat{k}_{AF})_{\kappa}\partial_{\mu}\phi\right] \notag \\
&\phantom{{}={}}+F_{\kappa\lambda}(\hat{k}_{\phi F})^{\kappa\lambda\mu}\partial_{\mu}\phi\,.
\end{align}%
The first term describes the kinematics of an eletromagnetism in (1+2) spacetime dimensions. The second is the kinematic term of the scalar field. The third and fourth are the direct successors of the \textit{CPT}-odd and \textit{CPT}-even modifications in (1+3) dimensions. The fifth is a Lorentz-violating contribution that involves the scalar field only. Note that this term has a form analogous to the $c$-type modifications of the (1+3)-dimensional scalar field theory introduced in the recent work~\cite{Edwards:2018lsn}. Hence, terms of an analog shape in (1+2) spacetime dimensions follow naturally from the electromagnetic SME sector in applying dimensional reduction. The $a$-type modifications of~\cite{Edwards:2018lsn} cannot be reproduced in this manner, though.

The remaining contributions describe mixings between the planar electromagnetic field and the scalar field. At the level of perturbation theory, these terms can be interpreted as vertices with an electromagnetic line and a scalar line meeting (see the perturbative treatment of the theory in \secref{sec:perturbative-expansion}). The interaction terms originating from the {\em CPT}-odd part comprise all components of the operator $(\hat{k}_{AF})_{\kappa}$, whereas only certain components of the {\em CPT}-even tensor $\hat{k}_F$ play a role in the interactions.

The structure of the modified Maxwell term remains untouched in the planar theory. The tensor $\hat{k}_F$ inherits its symmetries from the parent tensor. Taking these into account, there are 6 independent operators. Furthermore, getting rid of the unobservable double trace of~$\hat{k}_F$ by a redefinition of the electromagnetic fields, leads to the condition
\begin{equation}
\label{eq:double-tracelessness}
(\hat{k}_F)^{\mu\nu}_{\phantom{\mu\nu}\mu\nu}\overset{!}{=}0\,,
\end{equation}
reducing the number of independent operators to 5. They will be chosen as $(\hat{k}_F)^{0101}$, $(\hat{k}_F)^{0202}$, $(\hat{k}_F)^{0102}$, $(\hat{k}_F)^{0112}$, and $(\hat{k}_F)^{0212}$. The coefficient $(\hat{k}_F)^{1212}$ can be expressed by the first two of the previous five.

In contrast, the successor of the MCS term is quite different from its parent term. The vectorlike background field in two spatial dimensions boils down to an observer scalar $\hat{k}_{AF}$, which could simply be interpreted as a coupling constant similar to the topological mass in a Chern-Simons (CS) theory. For Lorentz violation of the minimal SME, the latter cannot involve preferred directions in spacetime. This finding agrees with the fact that there are genuine CS terms in an odd number of spacetime dimensions. These CS terms do not require vector or tensorlike background fields for their construction. It must be kept in mind, though, that in the nonminimal SME, $\hat{k}_{AF}$ contains preferred directions as well as four-derivatives. Last but not least, the observer four-vector~$(\hat{k}_{AF})_{\kappa}$ simply involves 3 independent coefficients in (1+2) dimensions and now plays the role of a coupling between the electromagnetic and the scalar field.

The Lorentz-violating term for the scalar field involves derivatives of this field. This makes sense, as the contribution originates from the modified Maxwell term. The scalar remains massless, as a mass term cannot be generated from the Lorentz-violating modifications.

The original operators $(\hat{k}_{AF})_{\hat{\kappa}}$ and $(\hat{k}_F)^{\hat{\kappa}\hat{\lambda}\hat{\mu}\hat{\nu}}$ are understood as infinite sums over sets of controlling coefficients suitably contracted with a number of four-derivatives successively increasing by 2. In this context it is important to recall that the mass dimensions of the fields $A^{\mu}$, $\phi$, and $F^{\mu\nu}$ change in a lower-dimensional spacetime. In particular,
\begin{equation}
[A^{\mu}]=\frac{1}{2}\,,\quad [\phi]=\frac{1}{2}\,,\quad [F^{\mu\nu}]=\frac{3}{2}\,.
\end{equation}
From these results we directly obtain the mass dimensions of the background fields:
\begin{equation}
[\hat{k}_F]=0\,,\quad [\hat{k}_{AF}]=1\,,\quad [\hat{k}_{\phi}]=0\,,\quad [\hat{k}_{\phi F}]=0\,.
\end{equation}
In principle, we can take over the SME notation for the controlling coefficients of mass dimension $d$, but we must keep in mind that $d$ no longer corresponds to the mass dimension of the field operator a controlling coefficient is contracted with. Nevertheless, the mass dimensions of the controlling coefficients follow the same rules as the parent coefficients. Since we will only work in momentum space later on, we express the decompositions of the background fields in terms of the four-momentum as follows:
\begin{subequations}
\label{eq:infinite-sums-lv-operators}
\begin{align}
(\hat{k}_{AF})_{\kappa}&=\sum_{d \text{ odd}}(k_{AF}^{(d)})_{\kappa}^{\phantom{\kappa}\alpha_{1}\dots \alpha_{(d-3)}}p_{\alpha_{1}}\dots p_{\alpha_{(d-3)}}\,, \displaybreak[0]\\[2ex]
(\hat{k}_F)^{\kappa\lambda\mu\nu}&=\sum_{d \text{ even}} (k_F^{(d)})^{\kappa\lambda\mu\nu\alpha_1\dots\alpha_{(d-4)}}p_{\alpha_{1}}\dots p_{\alpha_{(d-4)}}\,, \displaybreak[0]\\[2ex]
\hat{k}_{AF}&=\sum_{d \text{ odd}}(k_{AF}^{(d)})^{\alpha_{1}\dots\alpha_{(d-3)}}p_{\alpha_{1}}\dots p_{\alpha_{(d-3)}}\,, \displaybreak[0]\\[2ex]
(\hat{k}_{\phi})^{\kappa\mu}&=\sum_{d \text{ even}}(k_{\phi}^{(d)})^{\kappa\mu\alpha _{1}\dots\alpha_{(d-4)}}p_{\alpha_{1}}\dots p_{\alpha_{(d-4)}}\,, \displaybreak[0]\\[2ex]
(\hat{k}_{\phi F})^{\kappa\lambda\mu}&=\sum_{d \text{ even}}(k_{\phi F}^{(d)})^{\kappa\lambda\mu\alpha_{1}\dots\alpha_{(d-4)}}p_{\alpha_{1}}\dots p_{\alpha_{(d-4) }}\,,
\end{align}
\end{subequations}
where $p_{\mu}=\mathrm{i}\partial_{\mu}$. Each class of controlling coefficients has mass dimension $4-d$.

\section{Connection to physical fields}
\label{sec:connection-physical-fields}

Let us briefly review the concepts of electrodynamics in (1+2) dimensions \cite{Boito:2018rdh} that are most essential in the current context. The electromagnetic field strength tensor has 6 independent components in (1+3) dimensions, which amounts to 3 components for the electric field $\mathbf{E}$ and another 3 for the magnetic flux density $\mathbf{B}$. The situation in (1+2) dimensions is quite different, though. As an antisymmetric $(3\times 3)$ matrix has a maximum of 3 independent coefficients, the electric field in (1+2) dimensions is simply a vector with two components, $\mathbf{E}=(E^1,E^2)$, whereas the magnetic flux density is not even a vector at all, but is described by a scalar $B$. In an analog manner, the electric displacement field $\mathbf{D}$ is a vector with two components, $\mathbf{D}=(D^1,D^2)$, and the magnetic field strength $H$ is a simple scalar. Despite this mismatch of components, it is possible to write up the connection between $\mathbf{E}$, $B$ and $\mathbf{D}$, $H$ as a $(3\times 3)$ matrix equation as follows (compare to Eqs.~(4), (5) in~\cite{Kostelecky:2002hh}):
\begin{equation}
\begin{pmatrix}
\mathbf{D} \\
H \\
\end{pmatrix}=\begin{pmatrix}
\mathds{1}_2+\kappa_{DE}^{(2)} & \kappa_{DB}^{(2)} \\
\kappa_{HE}^{(2)} & 1+\kappa_{HB}^{(2)} \\
\end{pmatrix}\begin{pmatrix}
\mathbf{E} \\
B \\
\end{pmatrix}\,,
\end{equation}
where $\mathds{1}_2$ is the $(2\times 2)$ identity matrix. The counterparts of the quantities $\kappa_{DE}^{(2)}$, etc. in (1+3) dimensions are all $(3\times 3)$ matrices. Here, $\kappa_{DE}^{(2)}$ is a symmetric $(2\times 2)$ matrix with 3 components, $\kappa_{DB}^{(2)}$ is a two-component column vector, $\kappa_{HE}^{(2)}$ is a two-component row vector, and $\kappa_{HB}^{(2)}$ is a scalar:
\begin{subequations}
\begin{align}
\kappa_{DE}^{(2)ij}&\equiv -2(k_F)^{0i0j}\,, \\[2ex]
\kappa_{HB}^{(2)}&\equiv \frac{1}{2}\varepsilon^{pq}\varepsilon^{rs}(k_F)^{pqrs}\,, \\[2ex]
\kappa_{DB}^{(2)i}&\equiv \varepsilon^{pq}(k_F)^{0ipq}\,, \\[2ex]
\kappa_{HE}^{(2)i}&\equiv -(\kappa_{DB}^{(2)i})^T\,,
\end{align}
\end{subequations}
where $\varepsilon^{ij}$ denotes de Levi-Civita symbol in two (spatial) dimensions and $T$ stands for the transpose of a vector. The spatial indices simply run from 1 to 2. Now, the three components of $\kappa_{DE}^{(2)}$ and the two components of $\kappa_{DB}^{(2)}$ sum up to the five independent components of $k_F$. Note that $\kappa_{HB}^{(2)}$ does not deliver anything new due to the condition of \eqref{eq:double-tracelessness}, which translates to
\begin{equation}
\mathrm{Tr}\left(\kappa_{DE}^{(2)}+\frac{1}{2}\mathds{1}_2\kappa_{HB}^{(2)}\right)=0\,.
\end{equation}
Hence, in (1+2) dimensions, the number of electromagnetic phenomena is quite restricted. The matrix $\kappa_{DE}^{(2)}$ can still be interpreted as a permittivity tensor describing an optical medium with nontrivial refractive index. The connection between the fields $B$ and $H$ is a simple scaling factor that is made up of components of the previous permittivity tensor. Hence, there are no coefficients that are only tied to magnetic-field effects. The electric and magnetic field can still mix with each other and this mixing is described by two coefficients only. Furthermore, we can also define an isotropic coefficient in (1+2) dimensions as follows:
\begin{equation}
\kappa_{\mathrm{tr}}^{(2)}\equiv\frac{1}{2}\kappa_{DE}^{(2)ll}=-[(k_F)^{0101}+(k_F)^{0202}]=-(k_F)^{1212}\,,
\end{equation}
due to \eqref{eq:double-tracelessness}. Hence, this isotropic coefficient is not independent from the other coefficients, though --- in contrast to the situation in (1+3) dimensions. Finally, it is possible to construct objects $\tilde{\kappa}_{e+}$, etc. in the same manner as done in \cite{Kostelecky:2009zp}. As these definitions do not provide new insight, we will omit them here. It shall be mentioned that $\tilde{\kappa}_{e+}$, $\tilde{\kappa}_{e-}$ are $(2\times 2)$ matrices, whereas $\tilde{\kappa}_{o+}$ and $\tilde{\kappa}_{o-}$ are two-component column vectors. We found that $\tilde{\kappa}_{o-}=\mathbf{0}$, whereas all other objects involve nonzero components exclusively.

\section{Field equations}
\label{sec:field-equations}

We intend to derive the field equations from the Lagrange density (\ref{eq:lagrange-density-dimensionally-reduced}). As the theory involves higher derivatives, the Euler-Lagrange equations must be adapted accordingly. Based on the principle of least action, the Euler-Lagrange equations\footnote{See \cite{Woodard:2015zca} for a treatment in the context of classical mechanics, and \cite{Bollini:1986am} for field theories of higher derivatives.} for a field theory including up to the $n$-th derivative of a generic field $\psi$ read
\begin{align}
0&=\frac{\partial \mathcal{L}}{\partial\psi}-\partial _{\mu}\left(\frac{\partial\mathcal{L}}{\partial(\partial_{\mu}\psi)}\right)+\partial_{\mu}\partial_{\nu}\left(\frac{\partial\mathcal{L}}{\partial(\partial_{\mu}\partial_{\nu}\psi)}\right) \notag \\
&\phantom{{}={}}-\dots+(-1)^{n}\partial_{\mu_{1}}\dots\partial_{\mu_{n}}\left(\frac{\partial\mathcal{L}}{\partial(\partial_{\mu_{1}}\dots\partial_{\mu_{n}}\psi)}\right)\,.
\end{align}
Computing the appropriate derivatives, the field equations for the electromagnetic and the scalar field are given by:
\begin{subequations}
\label{eq:field-equations}
\begin{align}
\label{eq:field-equation-phi}
0&=\square\phi-2(\hat{k}_{\phi})^{\kappa\mu}\partial_{k}\partial_{\mu}\phi-\varepsilon^{\kappa\mu\nu}(\hat{k}_{AF})_{\kappa}F_{\mu\nu} \notag \\
&\phantom{{}={}}+(\hat{k}_{\phi F})^{\mu\kappa\lambda}\partial_{\mu}F_{\kappa\lambda}\,, \displaybreak[0]\\[2ex]
\label{eq:field-equation-photon}
0&=\partial_{\nu}F^{\mu\nu}-\varepsilon^{\mu\nu\rho}(\hat{k}_{AF})F_{\nu\rho}+(\hat{k}_{F})^{\mu\sigma\nu\rho}\partial_{\sigma}F_{\nu\rho} \notag \\
&\phantom{{}={}}-2\varepsilon^{\mu\nu\rho}(\hat{k}_{AF})_{\nu}\partial_{\rho}\phi+2(\hat{k}_{\phi F})^{\mu\nu\rho}\partial_{\nu}\partial_{\rho}\phi\,,
\end{align}
\end{subequations}
with the d'Alembertian $\square\equiv\partial_{\mu}\partial^{\mu}$. For zero Lorentz violation, the field equations reduce to the standard equation for a massless scalar field and the inhomogenous Maxwell equations without external sources. The field equation of the scalar field involves two kinetic terms, whereas there are three kinetic terms for the electromagnetic field due to the two distinct classes of operators.

Moreover, Eqs.~(\ref{eq:field-equations}) are coupled partial differential equations of at least second order (for minimal Lorentz violation). Higher than second derivatives can appear for nonminimal contributions. The solutions of the uncoupled scalar and electromagnetic field equations for external inhomogeneities (a conserved current in the electromagnetic case) will be determined later after deriving the corresponding Green's functions.

The components of the field strength tensor in two spatial dimensions have the following form:
\begin{equation}
\label{eq:components-field-strength-tensor}
F^{0i}=-E^i\,,\quad  F^{ij}=-\varepsilon^{ij}B\,.
\end{equation}
Hence, we see again that the electric field has two components, whereas the magnetic flux density is a simple scalar. The homogeneous Maxwell equations are Bianchi identities that are not modified by Lorentz violation. In contrast to the setting of (1+3) dimensions, there is only a single set of homogeneous Maxwell equations given by the Faraday law
\begin{equation}
\label{eq:faraday-law}
\varepsilon^{ij}\partial_iE^j=-\dot{B}\,.
\end{equation}
Now we investigate the inhomogeneous Maxwell equations (without external charge and current densities). They emerge as different components of the field equation (\ref{eq:field-equation-photon}) for vanishing couplings. First of all, we will keep the couplings, though. The zeroth component of (\ref{eq:field-equation-photon}) provides
\begin{align}
\label{eq:modified-gauss-law-position-space}
0&=\partial_{\nu}F^{0\nu}-\varepsilon^{0\nu\varrho}(\hat{k}_{AF})F_{\nu\varrho}+(\hat{k}_F)^{0\sigma\nu\varrho}\partial_{\sigma}F_{\nu\varrho} \notag \\
&\phantom{{}={}}-2\varepsilon^{0\nu\varrho}(\hat{k}_{AF})_{\nu}\partial_{\varrho}\phi+2(\hat{k}_{\phi F})^{0\nu\varrho}\partial_{\nu}\partial_{\varrho}\phi \notag \displaybreak[0]\\
&=-\partial_iE^i+2
(\hat{k}_{AF})B+2(\hat{k}_F)^{0i0j}\partial_iE^j-2(\hat{k}_F)^{0i12}\partial_iB \notag \\
&\phantom{{}={}}-2\varepsilon^{ij}(\hat{k}_{AF})_i\partial_j\phi+2(\hat{k}_{\phi F})^{0i0}\partial_i\dot{\phi}+2(\hat{k}_{\phi F})^{0ij}\partial_i\partial_j\phi\,,
\end{align}
where a dot on top of a field denotes a single time derivative. The spatial components lead to
\begin{align}
\label{eq:modified-ampere-law-position-space}
0&=\partial_{\nu}F^{i\nu}-\varepsilon^{i\nu\varrho}(\hat{k}_{AF})F_{\nu\varrho}+(\hat{k}_F)^{i\sigma\nu\varrho}\partial_{\sigma}F_{\nu\varrho} \notag \\
&\phantom{{}={}}-2\varepsilon^{i\nu\varrho}(\hat{k}_{AF})_{\nu}\partial_{\varrho}\phi+2(\hat{k}_{\phi F})^{i\nu\varrho}\partial_{\nu}\partial_{\varrho}\phi \notag \displaybreak[0]\\
&=\dot{E}^i+\varepsilon^{ij}(-\partial_jB+2(\hat{k}_{AF})E^j)-2(\hat{k}_F)^{0i0j}\dot{E}^j \notag \\
&\phantom{{}={}}+2(\hat{k}_F)^{0iij}(\varepsilon^{ij}\dot{B}+\partial_jE^i)+2(\hat{k}_F)^{0jij}\partial_jE^j \notag \\
&\phantom{{}={}}-2(\hat{k}_F)^{ijij}\varepsilon^{ij}\partial_jB+2\varepsilon^{ij}[(\hat{k}_{AF})_0\partial_j\phi-(\hat{k}_{AF})_j\dot{\phi}] \notag \\
&\phantom{{}={}}+2(\hat{k}_{\phi F})^{i00}\ddot{\phi}+2[(\hat{k}_{\phi F})^{ij0}-(\hat{k}_{\phi F})^{0ij}]\partial_j\dot{\phi} \notag \\
&\phantom{{}={}}+2(\hat{k}_{\phi F})^{iji}\partial_j\partial_i\phi+2(\hat{k}_{\phi F})^{ijj}\partial_j^2\phi\,.
\end{align}
Here, the physical fields were introduced via \eqref{eq:components-field-strength-tensor}. Note that the index $i$ in \eqref{eq:modified-ampere-law-position-space} is not summed over -- also when it appears twice or more often in a single term. From \eqref{eq:modified-gauss-law-position-space}, we obtain a modified Gauss law by setting the couplings to the scalar field equal to zero. In momentum space, the latter reads
\begin{equation}
\label{eq:modified-gauss-law}
0=p_iE^i-2\mathrm{i}(\hat{k}_{AF})B-2(\hat{k}_F)^{0i0j}p_iE^j+2(\hat{k}_F)^{0i12}p_iB\,.
\end{equation}
All Lorentz-violating operators are understood to be transformed to momentum space, as well, which corresponds to replacing all additional four-derivatives by four-momenta via $\partial_{\mu}=-\mathrm{i}p_{\mu}$ (cf.~the operators of Eqs.~(\ref{eq:infinite-sums-lv-operators})). Furthermore, from \eqref{eq:modified-ampere-law-position-space} we also obtain a modified Amp\`{e}re law of the form
\begin{align}
0&=p_0E^i-\varepsilon^{ij}(p_jB-2\mathrm{i}(\hat{k}_{AF})E^j)-2(\hat{k}_F)^{0i0j}p_0E^j \notag \\
&\phantom{{}={}}+2(\hat{k}_F)^{0iij}(\varepsilon^{ij}p_0B+p_jE^i)+2(\hat{k}_F)^{0jij}p_jE^j \notag \\
&\phantom{{}={}}-2(\hat{k}_F)^{ijij}\varepsilon^{ij}p_jB\,.
\end{align}
It is possible to eliminate the magnetic field from the latter by inserting the Faraday law (\ref{eq:faraday-law}) transformed to momentum space: $\varepsilon^{ij}p^iE^j=p_0B$. Doing so, leads to
\begin{align}
\label{eq:modified-ampere-law}
0&=\left[E^i-2(\hat{k}_F)^{0i0j}E^j\right]p_0^2 \notag \\
&\phantom{{}={}}+\left[p^j+2p_0(\hat{k}_F)^{0iij}+2p^j(\hat{k}_F)^{ijij}\right]\varepsilon^{ij}\varepsilon^{mn}p^mE^n \notag \\
&\phantom{{}={}}+2\mathrm{i}(\hat{k}_{AF})\varepsilon^{ij}p_0E^j-2\left[(\hat{k}_F)^{0iij}E^i+(\hat{k}_F)^{0jij}E^j\right]p_0p^j\,.
\end{align}
It can be checked quickly that a contraction of the modified Amp\`{e}re law (\ref{eq:modified-ampere-law}) with $p^i$ is equivalent to the modified Gauss law (\ref{eq:modified-gauss-law}) multiplied with $p_0$ after inserting the Faraday law.
The modified Amp\`{e}re law can also be expressed in matrix form as follows:
\begin{subequations}
\label{eq:modified-ampere-law-2}
\begin{align}
0&=\hat{N}^{ab}E^b\,, \displaybreak[0]\\[2ex]
\label{eq:determinant-modified-ampere-law}
\hat{N}^{ab}&=\left[\delta^{ab}-2(\hat{k}_F)^{0a0b}\right]p_0^2+2\mathrm{i}(\hat{k}_{AF})\varepsilon^{ab}p_0 \notag \\
&\phantom{{}={}}+\left[p^j+2p_0(\hat{k}_F)^{0aaj}+2(\hat{k}_F)^{ajaj}p^j\right]\varepsilon^{aj}\varepsilon^{mb}p^m \notag \\
&\phantom{{}={}}-2\left[(\hat{k}_F)^{0aaj}\delta^{ab}+(\hat{k}_F)^{0jaj}\delta^{jb}\right]p_0p^j\,.
\end{align}
\end{subequations}
The latter has nontrivial solutions for those $p_0$ only that satisfy the condition that the determinant of the coefficient matrix $\hat{N}$ vanishes. This condition corresponds to the modified dispersion equation of the theory. We will come back to that point later.

\section{Green's functions and perturbative expansion}
\label{sec:perturbative-expansion}

Gauge invariance prohibits a perturbative treatment of the theory described by \eqref{eq:lagrange-density-dimensionally-reduced}. Therefore, we add a gauge-fixing term\footnote{Note that the choice of an appropriate gauge fixing condition in quantization is nontrivial when higher derivatives are present (see, e.g., \cite{Galvao:1986yq} for Podolsky's extension of electrodynamics). In the context of the current paper, this problem can be ignored, as we do not carry out an explicit quantization of the planar electrodynamics presented. The only concept of quantum physics used is the propagator.} with the gauge-fixing parameter $\xi$ to the latter. By doing so, we define a new gauge-fixed Lagrange density indicated by the superfix `gf':
\begin{equation}
\mathcal{L}_{(1+2)}^{\mathrm{gf}}\equiv \mathcal{L}_{(1+2)}-\frac{1}{2\xi}(\partial\cdot A)^2\,.
\end{equation}
We now perform suitable partial integrations of the Lagrange density (\ref{eq:lagrange-density-dimensionally-reduced}) to write each term as an operator sandwiched by fields:
\begin{align}
\mathcal{L}_{(1+2)}^{\mathrm{gf}}&=\frac{1}{2}A^{\mu}(\square g_{\mu\nu}-\partial_{\mu}\partial_{\nu})A^{\nu}-\frac{1}{2}\phi\square\phi \notag \\
&\phantom{{}={}}-\varepsilon^{\lambda\mu\nu}A_{\lambda}(\hat{k}_{AF})\partial_{\mu}A_{\nu}-A_{\lambda}(\hat{k}_{F})^{\lambda\kappa\mu\nu}\partial_{\kappa}\partial_{\mu}A_{\nu} \notag \\
&\phantom{{}={}}+\phi(\hat{k}_{\phi})^{\kappa\mu}\partial_{\kappa}\partial_{\mu}\phi+2A_{\lambda}(\hat{k}_{\phi F})^{\lambda\kappa\mu}\partial_{\kappa}\partial_{\mu}\phi\ \notag \\
&\phantom{{}={}}+\varepsilon^{\nu\kappa\mu}\left[\phi(\hat{k}_{AF})_{\kappa}\partial_{\mu}A_{\nu}-A_{\nu}(\hat{k}_{AF})_{\kappa}\partial_{\mu}\phi\right] \notag \\
&\phantom{{}={}}+\frac{1}{2\xi}A^{\mu}\partial_{\mu}\partial_{\nu}A^{\nu}\,.
\end{align}
Partial integration must be carried out twice for some terms to arrive at this result. Boundary terms can be neglected in Minkowski spacetime. We transform the new Lagrange density to momentum space where it can be written in a suggestive form as follows:
\begin{subequations}
\begin{equation}
\mathcal{L}_{(1+2)}^{\mathrm{gf}}=\frac{1}{2}(A,\phi)
\begin{pmatrix}
\hat{M} & \hat{U}-\mathrm{i}\hat{V} \\
(\hat{U}+\mathrm{i}\hat{V})^{T} & \hat{S} \\
\end{pmatrix}\begin{pmatrix}
A \\
\phi
\end{pmatrix}\,,
\end{equation}
in terms of the $(3\times 3)$ matrix
\begin{equation}
\label{eq:matrix-operator-M}
\hat{M}_{\mu\nu}=-p^2\Theta_{\mu\nu}+\hat{K}_{\mu\nu}+\mathrm{i}\hat{L}_{\mu\nu}-\frac{p^{2}}{\xi}\Omega^{\mu\nu}\,,
\end{equation}
the scalar
\begin{equation}
\label{eq:scalar-operator-S}
\hat{S}=p^2-\hat{D}\,,
\end{equation}
the projectors
\begin{align}
\label{eq:projectors}
\Theta_{\mu\nu}&\equiv \eta_{\mu\nu}-\Omega_{\mu\nu}\,, \displaybreak[0]\\[2ex]
\Omega_{\mu\nu}&\equiv\frac{p_{\mu}p_{\nu}}{p^2}\,,
\end{align}
and the Lorentz-violating operators
\begin{align}
\label{eq:operators}
\hat{K}_{\mu\nu}&\equiv 2(\hat{k}_{F})_{\mu\kappa\beta\nu}p^{\kappa}p^{\beta}\,, \displaybreak[0]\\[2ex]
\hat{L}_{\mu\nu}&\equiv 2(\hat{k}_{AF})\varepsilon_{\mu\beta\nu}p^{\beta}\,, \displaybreak[0]\\[2ex]
\hat{U}_{\mu}&\equiv 2(\hat{k}_{\phi F})_{\mu\kappa\beta}p^{\kappa}p^{\beta}\,, \displaybreak[0]\\[2ex]
\hat{V}_{\mu}&\equiv 2\varepsilon_{\mu\kappa\nu}(\hat{k}_{AF})^{\kappa}p^{\nu}\,, \displaybreak[0]\\[2ex]
\hat{D}&\equiv 2(\hat{k}_{\phi})^{\kappa\mu}p_{\kappa}p_{\mu}\,.
\end{align}
\end{subequations}
The operator $\hat{L}_{\mu\nu}$ is antisymmetric, whereas $\hat{K}_{\mu\nu}$ is symmetric. Now we would like to construct the tools necessary for a perturbative treatment of this theory. Disregarding the background fields $(\hat{k}_{AF})^{\kappa}$ and $\hat{k}_{\phi F}$ switches off the mixing between the planar electromagnetic field and the scalar field. After doing so, we derive the Green's function for the scalar and electromagnetic field, respectively. The Lagrange density for the scalar field reads
\begin{equation}
\mathcal{L}_{\phi}=\frac{1}{2}\phi\hat{S}\phi\,,
\end{equation}
with the scalar operator of \eqref{eq:scalar-operator-S}. The Green's function corresponds to the inverse of the latter operator whose result is readily obtained:
\begin{equation}
\label{eq:inverse-scalar}
\Delta_{\phi}=\frac{1}{p^{2}-\hat{D}}\,.
\end{equation}
The treatment of the planar electromagnetic field is a bit more involved. We consider the Lagrange density
\begin{equation}
\label{eq:planar-photon-theory}
\mathcal{L}_A=\frac{1}{2}A_{\mu}\hat{M}^{\mu\nu}A_{\nu}\,.
\end{equation}
The inverse of the $(3\times 3)$ matrix $\hat{M}^{\mu\nu}$ can be expressed in terms of the metric tensor and suitable contractions of the original matrix. We found
\begin{subequations}
\label{eq:inverse-electrodynamics}
\begin{align}
\Delta_{\mu\nu}&=\frac{1}{\mathcal{R}}\left\{\frac{1}{2}\left[(\hat{M}_{\phantom{\alpha}\alpha}^{\alpha})^{2}-\hat{M}^{\alpha\beta}\hat{M}_{\beta\alpha}\right]\eta_{\mu\nu}\right. \notag \\
&\phantom{{}={}}\hspace{0.6cm}\left.-\,(\hat{M}_{\phantom{\alpha}\alpha}^{\alpha})\hat{M}_{\mu\nu}+\hat{M}_{\mu\beta}\hat{M}_{\phantom{\beta}\nu}^{\beta}\right\}\,,
\end{align}%
where the denominator $\mathcal{R}$ corresponds to
\begin{equation}
\label{eq:inverse-electrodynamics-denominator}
3!\mathcal{R}=(\hat{M}_{\phantom{\alpha}\alpha}^{\alpha})^{3}-3(\hat{M}^{\alpha\beta}\hat{M}_{\beta\alpha})(\hat{M}_{\phantom{\gamma}\gamma}^{\gamma})+2\hat{M}^{\alpha\beta}\hat{M}_{\beta\gamma}\hat{M}_{\phantom{\gamma}\alpha}^{\gamma}\,.
\end{equation}
\end{subequations}
We arrived at this result by using the Cayley-Hamilton theorem of linear algebra adapted to a pseudo-Euclidean space. Details are presented in~\ref{sec:general-photon-propagator}. In principle, it is possible to generalize Eqs.~(\ref{eq:inverse-electrodynamics}) to (1+$D$)-dimensional Minkowski spacetime and even to curved spacetimes, which will be discussed in the appendix.

The two-tensors that appear in Eqs.~(\ref{eq:inverse-electrodynamics}) can be totally formulated in terms of observer Lorentz scalars formed of the projectors of \eqref{eq:projectors} and observer two-tensors of \eqref{eq:operators}:
\begin{subequations}
\begin{align}
\hat{M}_{\phantom{\alpha}\alpha}^{\alpha}&=-\left(2+\frac{1}{\xi}\right)p^{2}+\hat{K}_{\phantom{\alpha}\alpha}^{\alpha}\,, \displaybreak[0]\\[2ex]
\hat{M}^{\alpha\beta}\hat{M}_{\beta\alpha}&=\left(2+\frac{1}{\xi^2}\right)p^4-2p^2\hat{K}_{\phantom{\alpha}\alpha}^{\alpha}-\hat{L}^{\alpha\beta}\hat{L}_{\beta\alpha} \notag \\
&\phantom{{}={}}+\hat{K}^{\alpha\beta}\hat{K}_{\beta\alpha}\,, \displaybreak[0]\\[2ex]
\hat{M}^{\alpha\beta}\hat{M}_{\beta\gamma}\hat{M}_{\phantom{\gamma}\alpha}^{\gamma}&=
-\left(2+\frac{1}{\xi^{3}}\right)p^6+3p^{4}\hat{K}_{\phantom{\alpha}\alpha}^{\alpha}
 \notag \\
&\phantom{{}={}}+3p^{2}(\hat{L}^{\alpha\beta}\hat{L}_{\beta\alpha}-\hat{K}^{\alpha\beta}\hat{K}_{\beta\alpha}) \notag \\
&\phantom{{}={}}+3(\mathrm{i}\hat{L}^{\alpha\beta}\hat{K}_{\beta\gamma}\hat{K}_{\phantom{\gamma}\alpha}^{\gamma}-\hat{L}^{\alpha\beta}\hat{L}_{\beta\gamma}\hat{K}_{\phantom{\gamma}\alpha}^{\gamma}) \notag \\
&\phantom{{}={}}+\hat{K}^{\alpha\beta}\hat{K}_{\beta\gamma}\hat{K}_{\phantom{\gamma}\alpha}^{\gamma}-\mathrm{i}\hat{L}^{\alpha\beta}\hat{L}_{\beta\gamma}\hat{L}_{\phantom{\gamma}\alpha}^{\gamma}\,.
\end{align}
These observer scalars involve the controlling coefficients to first, second, and third order, respectively. Due to the tensor structure of $\Delta_{\mu\nu}$, another observer two-tensor is indispensable:
\begin{align}
\hat{M}_{\mu\beta}\hat{M}_{\phantom{\beta}\nu}^{\beta}&=p^{4}\left(\Theta_{\mu\nu}+\frac{1}{\xi^2}\Omega_{\mu\nu}\right)-2p^{2}(\hat{K}+\mathrm{i}\hat{L})_{\mu\nu} \notag \\
&\phantom{{}={}}+(\hat{K}+\mathrm{i}\hat{L})_{\mu\beta}(\hat{K}+\mathrm{i}\hat{L})_{\phantom{\beta}\nu}^{\beta}\,.
\end{align}
\end{subequations}
Now, the Feynman rules needed for a perturbative treatment of the planar electrodynamics are given by:
\begin{fmffile}{diagrams}
\begin{subequations}
\begin{align}
\begin{array}{c}
\includegraphics{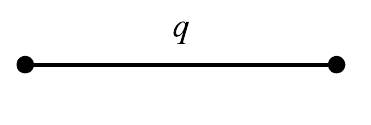}
\end{array}&=\mathrm{i}\Delta_{\phi}(q)\,, \displaybreak[0]\\[2ex]
\begin{array}{c}
\includegraphics{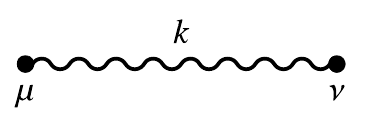}
\end{array}&=\mathrm{i}\Delta_{\mu\nu}(k)\,, \displaybreak[0]\\[2ex]
\begin{array}{c}
\includegraphics{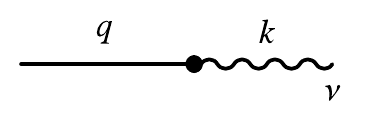}
\end{array}&=\begin{array}{l}
\mathrm{i}\varepsilon^{\nu\rho\sigma}\{[\hat{k}_{AF}(q)]_{\rho}q_{\sigma} \\
\hspace{1cm}-[\hat{k}_{AF}(k)]_{\rho}k_{\sigma}\} \\
-2[\hat{k}_{\phi F}(q)]^{\nu\varrho\sigma}q_{\varrho}q_{\sigma}\,. \\
\end{array}
\end{align}
\end{subequations}
\end{fmffile}
Here, $\Delta_{\phi}$ of \eqref{eq:inverse-scalar} and $\Delta_{\mu\nu}$ of \eqref{eq:inverse-electrodynamics} must be employed with the appropriate four-momentum inserted. The objects $\mathrm{i}\Delta_{\phi}$, $\mathrm{i}\Delta_{\mu\nu}$ correspond to the scalar and the electromagnetic propagator, respectively. The momentum directions at the vertex are understood as incoming. Several remarks are in order. First, for vanishing Lorentz violation, $\mathrm{i}\Delta_{\phi}$ corresponds to the propagator of a massless scalar field. Furthermore, $\mathrm{i}\Delta_{\mu\nu}$ reproduces the standard result for the electromagnetic propagator for our choice of the gauge-fixing term:
\begin{equation}
\label{eq:propagator-zero-lorentz-violation}
\mathrm{i}\Delta_{\mu\nu}\Big|_{\substack{k_F=0 \\ k_{AF}=0}}=\frac{-\mathrm{i}}{p^2}\left(\eta_{\mu\nu}-(1-\xi)\frac{p_{\mu}p_{\nu}}{p^2}\right)\,.
\end{equation}
For the equivalent of the 't Hooft-Feynman gauge in (1+2) dimensions, $\xi=1$, the second term simply does not contribute. The conventions have been chosen such that the standard propagators match the results stated in \cite{Peskin:1995}. Second, the denominator (\ref{eq:inverse-electrodynamics-denominator}) of $\Delta_{\mu\nu}$ is directly linked to the determinant of $\hat{M}$. The equation $\mathcal{R}=0$ can be interpreted as the dispersion equation of the modified planar electrodynamics. Therefore, its zeros with respect to $p_0$ are equal to the dispersion relations of planar electromagnetic waves. Third, the vertex involves the full set of planar $(\hat{k}_{AF})^{\kappa}$ operators, but only certain of the $\hat{k}_F$ (we already noticed this point before when discussing the interaction terms below \eqref{eq:lagrange-density-dimensionally-reduced}). The momentum dependencies of the operators have to be adapted appropriately to the momenta of the incoming scalar and electromagnetic field, respectively.

\section{Modified dispersion relations}
\label{sec:modified-dispersion-relations}

Asymptotic free plane-wave solutions are on-shell and satisfy some dispersion equation of the underlying theory. For the scalar field this means that $p^2-\hat{D}=0$, whereas electromagnetic waves satisfy the dispersion equation $\mathcal{R}=0$ with $\mathcal{R}$ given by \eqref{eq:inverse-electrodynamics-denominator}. Their zeros with respect to $p_0$ correspond to the dispersion relations of the sectors. We introduce the spatial momentum $\mathbf{p}=(p^1,p^2)$ that the dispersion relations are a function of. Let us first look at the scalar, which is easier to be treated. At leading order in the operator~$\hat{k}_{\phi}$, the positive-energy solutions are given by
\begin{subequations}
\label{eq:dispersion-relation-scalar-field}
\begin{align}
E^{(\pm)}(\mathbf{p})&=\frac{-2\hat{k}_{\phi}^{0i}p^i\pm\Psi(\hat{k}_{\phi})}{1-2\hat{k}_{\phi}^{00}}\bigg|_{p_0=\omega_0(\mathbf{p})}+\dots\,, \displaybreak[0]\\[2ex]
\Psi(\hat{k}_{\phi})&=\sqrt{4(\hat{k}_{\phi}^{0i}p^i)^2+(1-2\hat{k}_{\phi}^{00})(\mathbf{p}^2+2\hat{k}_{\phi}^{ij}p^ip^j)}\,,
\end{align}
\end{subequations}
where all additional $p_0$ are understood to be replaced by the standard massless dispersion relation $\omega_0(\mathbf{p})\equiv|\mathbf{p}|$. Thus, at leading order, there are two dispersion relations and they approach the standard result $\omega_0$ for vanishing controlling coefficients. When next-to-leading order effects in nonminimal frameworks are taken into consideration, there may be more than just two solutions. Furthermore, it is known that nonminimal theories provide dispersion relations that do not approach the standard result ($\omega_0$ in this case) for vanishing Lorentz violation. We will come back to this point below.

The next step is to discuss the dispersion relations of the modified, planar electromagnetic waves. They will be computed at leading order in Lorentz violation for the cases $\hat{k}_{AF}=0$ and $\hat{k}_F=0$, respectively. As $\hat{M}$ of the planar electromagnetic theory in \eqref{eq:planar-photon-theory} is a $(3\times 3)$ matrix where each coefficient involves at least two components of the four-momentum (for the minimal case), $\mathcal{R}$ is at least a polynomial of sixth degree in $p_0$. In what follows, we look at the positive dispersion relations, of which there are three for the minimal framework. Inserting the background fields, the denominator $\mathcal{R}$ can be written in the form
\begin{subequations}
\begin{align}
\label{eq:dispersion-equation-full}
\mathcal{R}&=-\frac{p^2}{\xi}\left\{p^2(p^2-\hat{K}^{\alpha}_{\phantom{\alpha}\alpha})-\frac{1}{2}\left[\hat{K}^{\alpha\beta}\hat{K}_{\beta\alpha}-(\hat{K}^{\alpha}_{\phantom{\alpha}\alpha})^2-\hat{L}^{\alpha\beta}\hat{L}_{\beta\alpha}\right]\right\} \notag \\
&=-\frac{p^4}{\xi}\mathcal{R}^{\text{phys}}\,,
\end{align}
with
\begin{align}
\label{eq:dispersion-equation-physical}
\mathcal{R}^{\text{phys}}&=p^2-\left(1+\frac{1}{2}(\hat{k}_F)^{\mu\nu}_{\phantom{\mu\nu}\mu\nu}\right)\hat{K}^{\alpha}_{\phantom{\alpha}\alpha} \notag \\
&\phantom{{}={}}+\hat{K}^{\mu\nu}(\hat{k}_F)_{\mu\kappa\nu}^{\phantom{\mu\kappa\nu}\kappa}-4(\hat{k}_{AF})^2\,.
\end{align}
\end{subequations}
The latter decomposition has been derived by inserting the explicit forms of $\hat{K}_{\mu\nu}$ and $\hat{L}_{\mu\nu}$. Using the condition of \eqref{eq:double-tracelessness}, we can omit the double trace of~$\hat{k}_F$. It is possible to extract the gauge fixing parameter in \eqref{eq:dispersion-equation-full} such that the physical dispersion relations do not depend on it. Furthermore, the second form of~$\mathcal{R}$ allows us to separate two powers of $p^2$ from the remaining expression $\mathcal{R}^{\text{phys}}$. Therefore, independently of the explicit choice of the Lorentz-violating background field, the standard dispersion relation in two spatial dimensions is a two-fold zero of $\mathcal{R}$ with respect to~$p_0$:
\begin{equation}
\label{eq:dispersion-relation-nonphysical}
\omega^{(1,2)}(\mathbf{p})=\omega_0\,.
\end{equation}
The third dispersion relation involves the Lorentz-violating operators. For the first case, we simply set $\hat{L}_{\mu\nu}=0$ and obtain:
\begin{subequations}
\label{eq:dispersion-relation-kF}
\begin{align}
\omega^{(3)}(\mathbf{p})|_{k_{AF}=0}&=\sqrt{\mathbf{p}^2+\frac{1}{2}\left[\hat{K}^{\alpha}_{\phantom{\alpha}\alpha}+\Upsilon(\hat{K})\right]}\Big|_{p_0=\omega_0(\mathbf{p})}+\dots\,, \displaybreak[0]\\[2ex]
\Upsilon(\hat{K})&=\sqrt{2\hat{K}^{\alpha\beta}\hat{K}_{\beta\alpha}-(\hat{K}^{\alpha}_{\phantom{\alpha}\alpha})^2}\,.
\end{align}
\end{subequations}
For the second case, we insert $\hat{K}_{\mu\nu}=0$, which leads to
\begin{equation}
\label{eq:dispersion-relation-kAF}
\omega^{(3)}(\mathbf{p})|_{k_F=0}=\sqrt{\mathbf{p}^2+4(\hat{k}_{AF})^2}\Big|_{p_0=\omega_0(\mathbf{p})}+\dots\,.
\end{equation}
Before we interpret these results, several remarks will be made. First, all $p_0$ components that are contracted with controlling coefficients on the right-hand sides of Eqs.~(\ref{eq:dispersion-relation-kF}), (\ref{eq:dispersion-relation-kAF}) are understood to be replaced by the standard dispersion relation $\omega_0$. Second, \eqref{eq:dispersion-relation-kF} involves the operator $\hat{k}_F$ at first order in Lorentz violation, whereas \eqref{eq:dispersion-relation-kAF} depends on $\hat{k}_{AF}$ only at second and higher orders. The latter property seems to be characteristic for (1+2) dimensions, as for (1+3) dimensions, $\hat{k}_{AF}$ enters the dispersion relation at first order \cite{Carroll:1989vb}.

Third, as there is only a single modified dispersion relation, birefringence does not occur. This property holds for the minimal theory,\footnote{cf.~\cite{Manojr:2011vd} where it was observed, as well} but also for the nonminimal framework at first order in Lorentz violation. For this reason, there are no coefficients associated with birefringence and it should be possible to parameterize the complete tensor structure of $\hat{k}_F$ by using an equivalent of the nonbirefringent \textit{Ansatz} of~\cite{Altschul:2006zz}. We found that
\begin{subequations}
\label{eq:nonbirefringent-ansatz}
\begin{align}
(\hat{k}_F)^{\mu\nu\varrho\sigma}&=\eta^{\mu\varrho}\hat{\tilde{k}}^{\nu\sigma}-\eta^{\mu\sigma}\hat{\tilde{k}}^{\nu\varrho}-\eta^{\nu\varrho}\hat{\tilde{k}}^{\mu\sigma}+\eta^{\nu\sigma}\hat{\tilde{k}}^{\mu\varrho}\,, \\[2ex]
\hat{\tilde{k}}^{\mu\nu}&\equiv (\hat{k}_F)_{\alpha}^{\phantom{\alpha}\mu\alpha\nu}\,,
\end{align}
\end{subequations}
where $(\hat{\tilde{k}}^{\mu\nu})$ is a symmetric and traceless $(3\times 3)$ matrix. The latter has 5 independent coefficients, whereupon \eqref{eq:nonbirefringent-ansatz} can represent the full tensor operator $\hat{k}_F$, indeed. In contrast to the counterpart of \eqref{eq:nonbirefringent-ansatz} in (1+3) dimensions, a global prefactor of 1/2 is not needed here.

Fourth, \eqref{eq:dispersion-relation-kF} has a form analog to one of the two dispersion relations obtained for the theory of minimal $(k_F)^{\mu\nu\varrho\sigma}$; cf.~Eqs.~(16) -- (18) in~\cite{Kostelecky:2002hh}. Last but not least, \eqref{eq:dispersion-relation-kAF} is the exact result for the minimal theory. This finding shows that the daughter theory is simpler than the parent theory, as the dispersion relation of MCS theory is more involved \cite{Carroll:1989vb}. Fifth, the negative-energy solutions found are related to the positive ones by reversing the sign of the spatial momentum components and the global sign in front of them. Let $\omega_-(\mathbf{p})$ be the negative-energy counterpart of a positive-energy dispersion relation $\omega_+(\mathbf{p})$. We then observe that $\omega_+(\mathbf{p})=-\omega_-(-\mathbf{p})$. This finding makes sense, as the photon is its own antiparticle. Note that the situation in the fermion sector is different and such correspondences do not necessarily hold for all sets of operators. The explanation is that the {\em C}-odd fermion coefficients have an opposite sign for antiparticles in comparison the particles.

An electromagnetic wave in (1+3) dimensions has two physical degrees of freedom and two nonphysical ones. The first correspond to the transverse polarizations and the second to the scalar and longitudinal one. The number of nonphysical degrees of freedom remains in (1+2) dimensions and the standard dispersion relation~(\ref{eq:dispersion-relation-nonphysical}) is associated with those. In (1+3) dimensions, the nonphysical dispersion relations can be eliminated by solving the modified Gauss law for one of the components of the electric field and eliminating this component from the modified Amp\`{e}re law (cf.~the second paper of \cite{Colladay:1996iz}). By doing so, the physical dispersion relations can be identified. This procedure does not seem to be necessary in (1+2) dimensions. The matrix $\hat{N}$ of the modified Amp\`{e}re law (\ref{eq:modified-ampere-law-2}) is given by
\begin{equation}
\det(\hat{N})=p_0^2\mathcal{R}_{\mathrm{phys}}\,,
\end{equation}
when \eqref{eq:double-tracelessness} is employed. Hence, the physical dispersion relations can be directly identified from the latter. The fact that the unphysical dispersion relations are not affected by Lorentz violation seems to be a property that holds in general (cf.~also~\cite{Schreck:2011ai}).

There is only a single polarization perpendicular to a given propagation dimension in two spatial dimensions. Therefore, for the minimal theory, there can be a single physical dispersion relation only. Further dispersion relations may emerge in nonminimal theories that include additional time derivatives. There are two classes of such dispersion relations. The first class only comprises perturbations of the standard dispersion relation. At leading order in Lorentz violation, these correspond to the dispersion relations~(\ref{eq:dispersion-relation-kF}), (\ref{eq:dispersion-relation-kAF}). The second class involves dispersion relations that do not approach the standard result for vanishing Lorentz violation (see, e.g., \cite{Ferreira:2019lpu,Schreck:2014fm} for examples in (1+3) dimensions). Such dispersion relations are sometimes called spurious. In principle, they are not necessarily unphysical, but they may be associated with ghosts. Besides, they are interpreted as Planck-scale effects, which means that they decouple from the theory for small momenta. For example, in~\cite{Ferreira:2019lpu} it was observed that such modes do not propagate in this regime, i.e., their group velocities go to zero. Explicit results must be determined on a case-by-case basis.

\section{Solutions to the free field equations}
\label{sec:solutions-free-field-equations}

At this point we have the necessary tools ready to deal with the field equations (\ref{eq:field-equations}). We consider the uncoupled equations that originate from the latter by setting the couplings equal to zero. Furthermore, we take inhomogeneities into account:
\begin{subequations}
\begin{align}
\label{eq:field-equation-scalar-free}
j(x)&=\square\phi-2(\hat{k}_{\phi})^{\kappa\mu}\partial_{\kappa}\partial_{\mu}\phi\,, \displaybreak[0]\\[2ex]
\label{eq:field-equation-photon-free}
j^{\mu}(x)&=\square A^{\mu}+\varepsilon^{\mu\nu\varrho}\hat{k}_{AF}F_{\nu\varrho}-(\hat{k}_F)^{\mu\sigma\nu\varrho}\partial_{\sigma}F_{\nu\varrho}\,,
\end{align}
\end{subequations}
where we used the Lorenz gauge condition $\partial\cdot A=0$ in the second equation to fix the gauge. The inhomogeneity associated with the scalar field is $j(x)$ and $j^{\mu}(x)$ is an external, conserved four-current density coupled to the electromagnetic field.

Let us first treat the scalar field. The general homogeneous solution is a superposition of plane-wave solutions involving the modified dispersion relations:
\begin{subequations}
\begin{align}
\phi^{\mathrm{hom}}(x)&=\int\frac{\mathrm{d}^2p}{(2\pi)^2}\sum_k\frac{1}{2E^{(k)}(\mathbf{p})}\phi^{(k)}(x)\,, \displaybreak[0]\\[2ex]
\phi^{(k)}(x)&=a^{(k)}(p)\exp(-\mathrm{i}p^{(k)}_{\alpha}x^{\alpha})+a^{(k)*}(p)\exp(\mathrm{i}p^{(k)}_{\alpha}x^{\alpha})\,.
\end{align}
\end{subequations}
Here, $a^{(k)}$ is an appropriate plane-wave amplitude, $a^{(k)*}$ its complex conjugate, and $(p^{(k)\alpha})=(E^{(k)},\mathbf{p})$ with the appropriate dispersion relations of Eqs.~(\ref{eq:dispersion-relation-scalar-field}). Note that all dispersion relations $E^{(k)}$ must be summed over. Spurious dispersion relations can, in principle, be omitted when the theory is restricted to its low-energy regime where the spurious modes decouple from the theory. The inhomogeneous solution of \eqref{eq:field-equation-scalar-free} can be written as a contour integral in the complex $p_0$ plane:
\begin{equation}
\phi^{\mathrm{in}}(x)=\frac{1}{(2\pi)^3}\int_{C_E}\mathrm{d}p_0\int\mathrm{d}^2p\,\Delta_{\phi}(p)\tilde{j}(p)\exp(-\mathrm{i}p_{\alpha}x^{\alpha})\,,
\end{equation}
with the Green's function $\Delta_{\phi}(p)$ of \eqref{eq:inverse-scalar}, the Fourier-transformed inhomogeneity $\tilde{j}(p)$, and an appropriate contour~$C_E$. By choosing retarded boundary conditions, the contour encircles all poles of $\Delta_{\phi}$ in counterclockwise direction.

The treatment of the electromagnetic field is a bit more challenging. The plane-wave homogeneous solutions of \eqref{eq:field-equation-photon-free} involve the modified polarization vectors as wave amplitudes:
\begin{subequations}
\begin{align}
A_{\mu}^{\mathrm{hom}}(x)&=\int\frac{\mathrm{d}^2p}{(2\pi)^2}\sum_k\frac{1}{2\omega^{(k)}(\mathbf{p})}A_{\mu}^{(k)}(x)\,, \displaybreak[0]\\[2ex]
A_{\mu}^{(k)}(x)&=\varepsilon_{\mu}^{(k)}(p)\exp(-\mathrm{i}p^{(k)}_{\alpha}x^{\alpha})+\varepsilon_{\mu}^{(k)*}(p)\exp(\mathrm{i}p^{(k)}_{\alpha}x^{\alpha})\,.
\end{align}
\end{subequations}
In the latter, $\varepsilon_{\mu}^{(k)}$ is the polarization vector of the $k$-th physical mode, $\varepsilon_{\mu}^{(k)*}$ its complex conjugated counterpart, and $(p^{(k)\alpha})=(\omega^{(k)},\mathbf{p})$ with the physical dispersion relation $\omega^{(k)}$. At leading order in Lorentz violation, these are given by Eqs.~(\ref{eq:dispersion-relation-kF}), (\ref{eq:dispersion-relation-kAF}). The polarization vectors $\varepsilon_{\mu}^{(k)}$ are basis solutions of \eqref{eq:field-equation-photon-free} in momentum space for $p_0=\omega^{(k)}$ and vanishing external four-current. These vectors are best obtained for particular cases, as their form critically depends on residual spacetime symmetries present. In the minimal framework, there is only a single physical polarization vector for this planar electrodynamics associated with the physical dispersion relation. The physical modes only must be summed over.

The inhomogeneous solution of \eqref{eq:field-equation-photon-free} for an external, conserved current density $j^{\mu}$ is obtained by means of the Green's function $\Delta_{\mu\nu}$ of \eqref{eq:inverse-electrodynamics}. Therefore, it can also be written as a contour integral in the complex $p_0$ plane:
\begin{equation}
\label{eq:inhomogeneous-solution-field-equations}
A_{\mu}^{\mathrm{in}}(x)=\frac{1}{(2\pi)^3}\int_{C_{\omega}}\mathrm{d}p_0\int\mathrm{d}^2p\,\Delta^{\text{phys}}_{\mu\nu}(p)\tilde{j}^{\nu}(p)\exp(-\mathrm{i}p_{\alpha}x^{\alpha})\,,
\end{equation}
where $\tilde{j}^{\mu}$ is the Fourier-transformed four-current density. Again, $C_{\omega}$ is an appropriate contour that encircles the physical poles in counterclockwise direction when retarded boundary conditions are chosen (cf.~\cite{Lehnert:2004be}). Physical solutions do not involve unphysical degrees of freedom, which is why they must be constructed from the physical part $\Delta^{\text{phys}}_{\mu\nu}$ of the Green's function. To get rid of the unphysical degrees of freedom in \eqref{eq:inverse-electrodynamics}, we decompose the latter into all observer two-tensors available. Therefore, we make the \textit{Ansatz}
\begin{subequations}
\begin{align}
\Delta_{\mu\nu}&=\alpha\eta_{\mu\nu}+\beta p_{\mu}p_{\nu}+\gamma(p_{\mu}\psi_{\nu}+\psi_{\mu}p_{\nu}) \notag \\
&\phantom{{}={}}+\delta(p_{\mu}\zeta_{\nu}+\zeta_{\mu}p_{\nu})+\epsilon\psi_{\mu}\psi_{\nu}+\iota\zeta_{\mu}\zeta_{\nu}+\kappa\varepsilon_{\mu\nu\alpha}p^{\alpha}\,,
\end{align}
where $\psi_{\mu}$ is a purely timelike and $\zeta_{\mu}$ a purely spacelike preferred direction:
\begin{equation}
(\psi^{\mu})=\begin{pmatrix}
1 \\
0 \\
0 \\
\end{pmatrix}\,,\quad (\zeta^{\mu})=\begin{pmatrix}
0 \\
1 \\
0 \\
\end{pmatrix}\,.
\end{equation}
\end{subequations}
The advantage of this form of the propagator is that it clearly separates the physical from the unphysical degrees of freedom. The unphysical ones are contained in terms that are proportional to at least one four-momentum with free Lorentz index. These terms vanish when the propagator is contracted with two conserved external four-currents due to $p\cdot \tilde{j}=0$. In this sense, the final term including the Levi-Civita symbol does not contribute to physics, as well. Hence, the physical degrees of freedom must be contained in the remaining terms only.

Setting the \textit{Ansatz} equal to the propagator leads to a system of equations for the parameters $\alpha \dots\kappa$ that can be solved with computer algebra. To make the calculation more feasible, we use the equivalent of Feynman-'t Hooft gauge in (1+2) dimensions, i.e., we set $\xi=1$. By doing so, the physical part of the propagator can be cast into the form
\begin{align}
\label{eq:propagator-physical}
\Delta_{\mu\nu}^{\mathrm{phys}}&=-\frac{1}{p^0p^1\mathcal{R}^{\text{phys}}}\left[(p^0p^1+\hat{K}^{01})\eta_{\mu\nu}+\left(\frac{p^0}{p^2}\hat{K}^{12}-\hat{K}^{01}\right)\psi_{\mu}\psi_{\nu}\right. \notag \\
&\phantom{{}={}}\hspace{1.9cm}\left.+\left(\hat{K}^{01}-\frac{p^1}{p^2}\hat{K}^{02}\right)\zeta_{\mu}\zeta_{\nu}\right]\,,
\end{align}
with the physical dispersion equation of \eqref{eq:dispersion-equation-physical}. An excellent cross check for the correctness of this result is that the denominator $p^4$ giving rise to unphysical dispersion relations cancels completely from $\Delta_{\mu\nu}^{\text{phys}}$, as expected. Besides, it was checked that
\begin{equation}
\tilde{j}^{\mu}(\Delta_{\mu\nu}-\Delta_{\mu\nu}^{\text{phys}})\tilde{j}^{\nu}=0\,.
\end{equation}
Thus, a contraction of $\Delta_{\mu\nu}$ with two external, conserved four-currents gets rid of all unphysical degrees of freedom, which is why the result must be the same as contracting $\Delta_{\mu\nu}^{\text{phys}}$ with these currents. Note that the same vector field must be obtained in \eqref{eq:inhomogeneous-solution-field-equations} by contracting $\Delta_{\mu\nu}$ with $\tilde{j}^{\mu}$ from the left or with $\tilde{j}^{\nu}$ from the right. Another observation is that the numerator of \eqref{eq:propagator-physical} involves Lorentz violation at first order only. This is also the reason why $\hat{k}_{AF}$ does not contribute to the numerator, but just to the denominator $\mathcal{R}^{\text{phys}}$.

As the unphysical poles have been eliminated, the contour~$C_{\omega}$ in the complex $p_0$ plane only encircles the positive physical poles and their negative counterparts in counterclockwise direction. The contour integral over $p_0$ is usually evaluated with the residue theorem. We will only consider the minimal case explicitly. Let $\omega^{(3)}$ be the physical dispersion relation. The denominator of \eqref{eq:propagator-physical} can then be written as
\begin{subequations}
\begin{align}
\mathcal{R}^{\text{phys}}&=\Xi[p^0-\omega^{(3)}(\mathbf{p})][p^0+\omega^{(3)}(-\mathbf{p})]\,, \\[2ex]
\Xi&=1-2\left[(\hat{k}_F)^{0101}+(\hat{k}_F)^{0202}\right] \notag \\
&\phantom{{}={}}+4\left\{(\hat{k}_F)^{0101}(\hat{k}_F)^{0202}-\left[(\hat{k}_F)^{0102}\right]^2\right\}\,.
\end{align}
\end{subequations}
The residues are readily obtained from this form of the denominator. We already mentioned that additional physical dispersion relations may arise for the nonminimal case. Those must be taken into account in the contour integral. Finally, in the limit of zero Lorentz violation, we obtain
\begin{equation}
\mathrm{i}\Delta_{\mu\nu}^{\text{phys}}\Big|_{\substack{k_F=0 \\ k_{AF}=0}}=\frac{-\mathrm{i}\eta_{\mu\nu}}{p^2}\,,
\end{equation}
as expected from \eqref{eq:propagator-zero-lorentz-violation}.

\section{Conclusions}
\label{sec:conclusions}

In the current paper, we constructed the framework of a Lorentz-violating extension of electrodynamics in (1+2) spacetime dimensions including field operators of arbitrary mass dimension. The latter was obtained from the electromagnetic sector of the nonminimal SME by applying the method of dimensional reduction. The resulting modified planar electrodynamics involves an additional scalar field that the electromagnetic field can mix with. We obtained the set of coupled field equations, the modified Maxwell equations, the Green's functions for the electromagnetic and the scalar field, and the perturbative Feynman rules of the theory. The modified dispersion relations for electromagnetic waves were computed at leading order in Lorentz violation. Finally, the general homogeneous solutions of the uncoupled field equations were constructed as well as the inhomogeneous solutions by means of the Green's functions. In the case of electromagnetic waves, the unphysical degrees of freedom had been removed from the Green's function before. More sophisticated analyses of the properties of particular interesting sectors of this framework are planned to be carried out in the future.

The framework developed can serve as a base for theoretical investigations of electromagnetic aspects in planar condensed-matter systems. Also, the structure of the planar electrodynamics is simpler than the parent theory in (1+3) spacetime dimensions. This enabled us to perform certain computations that would have been much more challenging for the parent theory such as the elimination of the unphysical degrees of freedom from the Green's function of the electromagnetic field. With this methodology and experience at hand, similar general results are at reach for the electromagnetic sector of the nonminimal SME.

\section*{Acknowledgments}

The authors thank the Brazilian research agencies CNPq, CAPES, and FAPEMA for financial support. In particular, M.M. Ferreira is grateful to FAPEMA Universal 00880/15; FAPEMA PRONEX 01452/14; CNPq Produtividade 308933/2015-0. M. Schreck is indebted to FAPEMA Universal 01149/17; CNPq Universal 421566/2016-7; CNPq Produtividade 312201/2018-4.

\appendix
\section{General propagator of modified photons in (1+$D$) dimensions}
\label{sec:general-photon-propagator}

Here we would like to outline the construction of \eqref{eq:inverse-electrodynamics}. Inverse matrices can be obtained in a systematic way via the Cayley-Hamilton theorem of linear algebra. Note that in the mathematics literature, the corresponding formulas are usually stated in Euclidean space. As we have not found any text that discusses how to adapt these formulas to a pseudo-Euclidean setting, we intend to present these results here for a general purpose.

We consider a free theory of a vector field $A_{\mu}$ in (1+$D$)-dimensional Minkowski spacetime described by the Lagrange density
\begin{equation}
\mathcal{L}_A=A_{\mu}\hat{\mathcal{H}}^{\mu\nu}A_{\nu}\,,
\end{equation}
where $\hat{\mathcal{H}}^{\mu\nu}$ is a tensor operator in momentum space that can contain arbitrary powers of the four-momentum. The Lorentz indices run from 0 to $D$ and they are contracted with each other via the (1+$D$)-dimensional Minkowski metric $\eta_{\mu\nu}$. The propagator can be obtained from the inverse of the operator $\hat{\mathcal{H}}^{\mu\nu}$ that is constructed in a general way as follows:
\begin{subequations}
\label{eq:general-inverse}
\begin{align}
\label{eq:general-inverse-Delta}
\Delta_{\mu\nu}&=\frac{\det(\eta_{\alpha\beta})}{\mathcal{R}}\sum_{k=0}^D (-1)^{D+k}\frac{\hat{\mathcal{H}}_{\mu\nu}^{D-k}}{k!}\Lambda_k\,, \displaybreak[0]\\[2ex]
\Lambda_k&=B_k(s_1,-s_2,2!s_3,\dots,(-1)^{k-1}(k-1)!s_k)\,,
\end{align}
\end{subequations}
where $B_k$ are the Bell polynomials. The latter are tabulated with the first four given by
\begin{subequations}
\begin{align}
B_0&=1\,, \displaybreak[0]\\[2ex]
B_1(x_1)&=x_1\,, \displaybreak[0]\\[2ex]
B_2(x_1,x_2)&=x_1^2+x_2\,, \displaybreak[0]\\[2ex]
B_3(x_1,x_2,x_3)&=x_1^3+3x_1x_2+x_3\,.
\end{align}
\end{subequations}
Hence, the Bell polynomial $B_k$ is a function of $k$ variables. The matrices $\hat{\mathcal{H}}_{\mu\nu}^k$ that appear in \eqref{eq:general-inverse} are contractions of $k$ matrices $\hat{\mathcal{H}}^{\mu\nu}$ such that the first index of the first matrix and the second of the last one remain free:
\begin{subequations}
\begin{align}
\hat{\mathcal{H}}_{\mu\nu}^{0}&=\mathcal{I}_{\mu\nu}\,, \displaybreak[0]\\[2ex]
\hat{\mathcal{H}}_{\mu\nu}^{1}&=\hat{\mathcal{H}}_{\mu\nu}\,, \displaybreak[0]\\[2ex]
\hat{\mathcal{H}}_{\mu\nu}^{2}&=\hat{\mathcal{H}}_{\mu\beta}\hat{\mathcal{H}}_{\phantom{\beta}\nu}^{\beta}\,, \displaybreak[0]\\[0ex]
&\phantom{{}={}}\vdots \notag \\[1ex]
\hat{\mathcal{H}}^{k}_{\mu\nu}&=\hat{\mathcal{H}}_{\mu\alpha_{1}}\hat{\mathcal{H}}^{\alpha_{1}\alpha_{2}}\hat{\mathcal{H}}_{\alpha_{2}\alpha_{3}}\dots\hat{\mathcal{H}}_{\phantom{\alpha_k}\nu}^{\alpha_{k}}\,.
\end{align}
The tensor components of $\mathcal{I}$ are defined via the inverse of the metric tensor:
\begin{equation}
\mathcal{I}_{\mu\nu}\equiv(\eta^{-1})_{\mu\nu}\,.
\end{equation}
\end{subequations}
Clearly, the inverse of the Minkowski metric reproduces this metric. However, we will work with the general definition to be able to generalize the results to arbitrary spacetimes.

The variables $s_k$ are given by traces of the latter contractions, i.e., $s_k=(\hat{\mathcal{H}}^k)^{\alpha}_{\phantom{\alpha}\alpha}$. The denominator $\mathcal{R}$ in front of the sum in \eqref{eq:general-inverse-Delta} corresponds to the determinant of the matrix $\hat{\mathcal{H}}^{\mu\nu}$ that can be expressed in the form
\begin{equation}
\label{eq:general-determinant}
\mathcal{R}=\frac{\det(\eta_{\mu\nu})}{(1+D)!}\Lambda_{D+1}\,.
\end{equation}
The characteristic polynomial that $\hat{\mathcal{H}}$ satisfies is given as
\begin{subequations}
\begin{equation}
p(\hat{\mathcal{H}})=\hat{\mathcal{H}}^{D+1}+c_{D}\hat{\mathcal{H}}^{D}+\ldots+c_{1}\hat{\mathcal{H}}+c_{0}\mathcal{I}=0\,,
\end{equation}
where
\begin{equation}
c_{D-k+1}=\frac{(-1)^{k}}{k!}\Lambda_k\,.
\end{equation}
\end{subequations}
The latter is the main ingredient used in the proof of the general formula (\ref{eq:general-inverse}) of the inverse.

It is worthwhile to notice that the formulas above have been expressed in a covariant form. Thus, this inverse does not lose its validity for an electromagnetic theory defined in a curved spacetime. In this case, all occurrences of the Minkowski metric $\eta_{\mu\nu}$ must be replaced by the corresponding pseudo-Riemannian metric $g_{\mu\nu}$ in (1+$D$) dimensions.


\end{document}